\documentclass[12pt]{article}
\usepackage{graphicx,psfrag,epsf}
\usepackage{enumerate}
\usepackage{url}
\usepackage{tabularx}
\usepackage{amsmath}
\usepackage{amsfonts}
\usepackage{amsthm}
\usepackage{amssymb}
\usepackage{multimedia}
\usepackage{verbatim}
\usepackage{times}
\usepackage[T1]{fontenc}
\usepackage{booktabs}
\usepackage{tabularx}
\usepackage{colortbl}
\usepackage{graphicx}
\usepackage{amsmath}
\usepackage{amsfonts}
\usepackage{amsthm}
\usepackage{framed}
\usepackage{natbib}
\usepackage{ulem}  
\usepackage{algorithmic}
\usepackage{algorithm}
\usepackage{subcaption}
\usepackage{caption}

\newcolumntype{L}[1]{>{\raggedright\arraybackslash}p{#1}} 
\newcolumntype{C}[1]{>{\centering\arraybackslash}p{#1}} 
\newcolumntype{R}[1]{>{\raggedleft\arraybackslash}p{#1}} 

\begin{document}

\def\spacingset#1{\renewcommand{\baselinestretch}%
{#1}\small\normalsize} \spacingset{1}


  \title{\bf Space Filling Split Plot Design using Fast Flexible Filling.}
  \author{Thomas Muehlenstaedt \thanks{
    The author gratefully acknowledges the support of W. L. Gore \& Associates and especially the members of its Global Statistics Team.},  Maria Lanzerath, \hspace{.2cm}\\
    Autonomous Intelligent Driving GmbH, W. L. Gore \& Associates\\}

 \maketitle

\begin{abstract}
In this article, an adaption of an algorithm for the creation of experimental designs by \cite{LekivetzJones2014FFF} is suggested, dealing with constraints around randomization. Split-plot design of experiments is used, when the levels of some factors cannot be modified as easily as others. While most split-plot designs deal in the context of I-optimal or D-optimal designs for continuous response outputs, a space filling design strategy is suggested in here. The proposed designs are evaluated based on different design criteria, as well as an analytical example. 
\end{abstract}

\vspace{0.5cm}

Keyworks: Design of Experiments, Blocking, Uncertainty Quantification 
\section{Introduction}

This work is motivated by a project at W. L. Gore \& Associates: a number of different simultaneous requirements created a novel design situation, as described in section 3. In order to meet those requirements, a small adaption of an algorithm by \cite{LekivetzJones2014FFF} is made. It incorporates a blocking structure into a space filling design using hierarchical clustering approaches. Therefor, a motivation is given in the following section. Requirements and possible statistical models are discussed on section 3 and 4, and the actual algorithm is described in section 5. As we are convinced of the broader applicability of these designs, section 6 describes potential different use cases.  Finally, the suggested designs are compared to some others based on optimality criteria, and the analysis of two sample data sets based on the new design type is performed.

\section{Application example and motivation}

Due to confidentiality reasons, the use case can only be described on a high level: In a manufacturing process, 2 responses are measured from the produced items. One is a binomial output ($Y_1$, pass/fail), and the other one is a continuous output ($Y_2$), which is to be maximized.
The two outputs are competing: generally, if $Y_1$ is optimized to pass, $Y_2$ will be lowered, and vice versa.
The experiment involved 5 continuous variables, i.e. a speed, two different temperatures and two different air flow rates. Additionally, two  products are made with this process, though the experiment deals with just on one of them. 
In a screening experiment we learned there is very little experimental or measurement error, the replicates showed a very high reproducibility for both outputs with no measurable random noise.

In a next step, an optimization experiment for that process is designed to find a factor setting that creates a pass result for $Y_1$, while the $Y_2$ output gets as high as possible.
The experiment requires a split-plot structure, because the two temperatures are hard to change. This is because heating up the system requires a waiting time to reach a constant level and, as there is no active cooling system, cooling down the temperature requires even longer waiting times. The two temperatures become so called \textit{hard to change factors}.
Furthermore, some combinations of temperatures and air flows can be omitted from the design: they are known to deliver either a $Y_1 = \text{fail}$, or almost certainly a too low $Y_2$ value. This situation is called a \textit{constrained design space}.

\section{Requirements for a solution}
Translating the above example into a list of requirements, the following can be listed:
\begin{enumerate}
	\item Hard to change factors will only be varied in blocks.
	\item The design has good space filling properties in its whole plot variables, subplot variables, and combinations of them.
	\item The design allows good fitting for both the binomial and the continuous (normally distributed) output.
	\item If possible, allow for constrained design space options.
	\item If possible, allow for categorical factors to be incorporated.
\end{enumerate}

 The main approach to fulfill the requirements 1 through 3 is an adaption of the strategy described in \cite{LekivetzJones2014FFF}. It is the utilization of hierarchical cluster algorithms for the design creation. \\
 Points 4 and 5 will not be treated at in the paper, however, we want to emphasize this approach can be extended to those two requirements. 

\section{Models}

Traditional split plot experiments are analyzed by a linear mixed model as in the following equation: 
\begin{equation}
Y = X \beta +  e_{WP} + e_{SP},
\end{equation}
(\cite{Goos2002optBlockedSplitPlotDoEBook}), where $Y$ is a continuous measurement, and $X$ is a regression matrix representing e.g. a purely linear model, or a quadratic model linear in the coefficient vector $\beta$. In the referenced example, the output could be sufficiently represented by a quadratic model. However, if a process shows more complex behavior, a quadratic model might not be sufficient, and a higher order polynomial model would provide a more adequate fit. In the motivating example, no measurable error existed, and data were reproducible up to the precision of the measurement device. Hence, the 2 error terms do not represent actual behavior here. 

In the case of a very high signal to noise ratio (up to no measurable error), models steaming from geoscience are an attractive choice, as they are often made for deterministic data as well as for complex model outputs. Especially gaussian process modeling (aka Kriging) has gained much attention in several application areas. With the model equation for Kriging being quite similar at first hand sight:
\begin{equation}
Y = X \beta + e_{spatial},
\end{equation}
the big difference comes through the included parametric correlation structure (\cite{SanWilNot03desi}, \cite{Cressie1993Geostats}), based on distances of the input factor points: 

\begin{equation}
cov(e_{spatial}(x_1),e_{spatial}(x_2)) = \sigma_{spatial}^2 k_{\theta}(x_1 - x_2), 
\end{equation}
with $k_{\theta}(.)$ being a kernel function. This correlation structure enables the model to fit quite complex behaviors as well as to exactly reproduce observed data points. In general, there are also adaptions of these kinds of models for categorical output (\cite{DiggleRibeiro2007ModelBasedGeostats}), yet they are not considered here, as no implementation for the considered examples in R is known to the authors. 

Due to the categorical nature of one output, the model needs to be adapted. A generalized linear model version of the above equation is chosen:
\begin{equation}
E(Y) = g(X \beta + e),
\end{equation}
it also allows to include an additional error term for running an experiment in blocks (\cite{McCullaghNelder1989GLM}).

Two alternative general purpose prediction models could be considered. Those are neural networks (\cite{Goodfellow2016DL}) and support vector machines (\cite{Schoelkopf2018SVM}). Both can handle continuous and categorical output. However, these methods are more frequently used in data mining situations with large data sets and less in situations of designed experiments. For neural networks, the main feature is to stack layers of linear combinations, connected by nonlinear activation functions. Finding suitable parameters is done via optimization of a loss function, typically with a variant of gradient descent algorithms. Support vector machines are usually used as a so called kernel method.

\section{DoE Algorithm}

As mentioned earlier, the main reference and motivation for the procedure below is an article by \cite{LekivetzJones2014FFF}. They use hierarchical Ward clustering algorithms to generate space filling designs. Hierarchical clustering algorithms are explained for example in \cite{Everit2001ClusterInto}, and evaluated in \cite{Murtagh1983SurveyHierarchicalClustering}.

The following details serve as an input for the algorithm: 
\begin{itemize}
	\item Overall number of experiments ($n_{overall}$)
	\item Number of whole plots ($n_{WP} << n_{overall}$)
	\item Number of whole plot factors (hard to change), $(d_{WP})$ and number of subplot factors (easy to change) $(d_{SP})$ (hence also the overall number of factors $d$)
	\item Number of random points generated for the clustering algorithm to start from ($n_{sim} >> n_{overall}$).
\end{itemize}
The design space is always assumed to be $[-1,1]^d$.
As for the original fast flexible filling (FFF) design, the Ward clustering algorithm is the main work horse here as well.

To adopt the idea from \cite{LekivetzJones2014FFF}, the clustering is done in 2 steps. The first step is the same as for the original FFF algorithm. That is, a standard Ward clustering is done until the randomly initialized points $X ~ (n_{sim}, d)$ are combined into $n_{overall}$ clusters using Ward's rule for combining clusters: If $\bar{x}_k$ and $\bar{x}_l$ are two cluster averages with cluster sizes $N_k$ and $N_l$ among a number of clusters $n_{cl}$, then combine the two clusters into one, which minimizes
\begin{equation}
D_{kl} = \frac{\| \bar{x}_k - \bar{x}_l\|^2}{1/N_k + 1/N_l}.
\end{equation}

In the second step, only the $d_{WP}$ hard to change factors are used to continue the Ward clustering.  Basically, the clustering starts again using only the $d_{WP}$ factors of the $n_{overall}$ runs. This is done until the number of clusters for the whole plot factors reaches $n_{WP}$. In this stage it is important to track the whole plot cluster that each of the $n_{overall}$ runs belongs to. In the second part of the algorithm, clusters, cluster averages, and cluster distances carry the superscript $(.)^{WP}: C^{WP}_{l}, D^{WP}_{kl}, \bar{x}^{WP}_{l} $. This is to make clear they only relate to the $WP$ factors.

In the second part of the algorithm, clusters, cluster averages, and cluster distances only relate to the whole plot factors, hence indicated by $.^{WP}: C^{WP}_{l}, D^{WP}_{kl}, \bar{x}^{WP}_{l} $.\\

One main advantage of the Ward clustering algorithm is seen in its recursive formula for identifying the two clusters to be joined. This can be described as joining the two clusters, where the Euclidean distance between the cluster means is minimal.

\begin{algorithm} 
	\caption{Split Plot FFF Algorithm} 
	\label{alg1} 
	\begin{algorithmic} 
		\REQUIRE $n_{sim} >> n_{overall} > n_{WP} > 0$, $d, d_{SP}, d_{WP} \in \mathbb{N}, d = d_{WP} + d_{SP} $
		\STATE Generate $n_{sim}$ uniform random numbers $x_i, i = 1, \dots, n_{sim}$
		\STATE Define $C_i = \{x_i\}, i = 1, \dots, n_{sim}$
		\WHILE{$n_{sim} > n_{overall}$}
			\STATE Search pair $k, l$ with $D_{kl}$ minimal according to eq. (5)
			\STATE Assign all points in $C_l$ and $C_k$ to joint 
			cluster  $C_l := \{x: x\in C_l \lor x \in C_k\}$
			\STATE Update Cluster Average $\bar{x}_{l} = \frac{1}{|C_l|}\sum_{x \in C_l} x$
			\STATE $n_{sim} := n_{sim} - 1$
		\ENDWHILE		
		\WHILE{$n_{sim} > n_{WP}$}
			\STATE Search pair $k, l$ with $D^{WP}_{kl}$ minimal according to eq. (5)
			\STATE Assign all points in $C^{WP}_l$ and $C^{WP}_k$ to joint 
			cluster  $C^{WP}_l := \{x^{WP}: x^{WP} \in C^{WP}_l \lor x^{WP} \in C^{WP}_k\}$
			\STATE Update Whole Plot Cluster Average $\bar{x}^{WP}_{l} = \frac{1}{|C_l|}\sum_{x^{WP} \in C_l} x^{WP}$
			\STATE $n_{sim} := n_{sim} - 1$
		\ENDWHILE	
		\STATE Report $n_{overall}$ cluster averages $\bar{x}_i = [\bar{x}_i^{WP}, \bar{x}_i^{SP}]$, and for the whole plot factors, a column with their cluster assignments to the whole plots.
	\end{algorithmic}
\end{algorithm}

\section{Possible applications}

While this situation may seem very specific, these designs can be applied in a number of situations: the motivating example shows an application in a manufacturing process with a high signal to noise ratio, and the need to treat factors differently in their randomization scheme. This often occurs in industrial applications when e.g. a temperature is a design factor, and its cooling and heating times reveal as time consuming. Also, when reassembling a machine is required, a split plot structure might make sense. 

Simulation experiments can be another area of application, e.g. continuous fluid dynamics simulations, or other modeling approaches based on physics. Although most of the factors can be varied easily, resetting a factor sometimes holds challenges. This can be the case if there is no automatic process to change the underlying meshing, or if there is a chain of different simulations with no automatic way to couple these. 

A third application is seen in the context of machine learning hyper parameter optimization (\cite{Shahriari2016TakingHumanOutOfTheLoop}). Baysian methods are used to optimize parameters like learning rates, optimization algorithms, or image resolution, e.g. for computer vision tasks. Parameters like the image resolution may require a large stack of images to be pre-processed and stored at separate locations for each run. Taking parameters like the image resolution as hard to change would facilitate the training of a larger amount of models, and would also facilitate paralleling training jobs.

\section{DoE comparison}
In the following, a simulation study is conducted in order to compare the suggested DoE's to other design options: in a first step, the different designs are compared using optimality criteria. In a second step, a subset of the designs gets compared on a test function regarding their predictive power. The expectation is not to outperform other DoEs but to observe no large drop in performance.
\subsection{Design options for comparison}
The following DoE's are considered in the comparison: 
\begin{itemize}
	\item A Split Plot RSM Design (Split Plot I optimal) constructed by JMP's Custom Design platform (\cite{JMP15}),
	\item A space filling Split Plot Design (Split Plot FFF) using a fast flexible filling strategy as described above,
	\item A standard Fast Flexible Filling Design without Split Plot Structure (FFF), also created in JMP,
	\item A maximin Latinhypercube (LHS) as described in the R-package lhs (\cite{R_lhs}).
\end{itemize}
Other options would potentially be Baysian D-optimal designs as implemented in the R-package acebayes (\cite{R_acebayes}). However, the package acebayes was not functional when applied by the authors. For the comparison, the following parameters are chosen: 
\begin{itemize}
	\item No. of factors: 2 and 4 with 1 or 2 whole plot factors respectively for the designs capable of providing a split plot structure.
	\item No. of runs: Between 20 and 50, in steps of 5.  
	\item For I-optimal designs and SPFFF designs, number of whole plots is varied within $\{\{8, 12, 16\}$
	\item All possible combinations of the ranges above are evaluated. Only expection is for $n = 20$, no design with 16 whole plots is created.
\end{itemize}

\subsection{Comparison by design criteria}

We recommend to apply a variety of criteria to evaluate several performance aspects of a design. Those criteria can be put in two different classes, one for space filling designs, and one for classical statistical models. We apply 5 criteria in this article: the maximin criterium, a variant of maximin developed by \cite{MorMit95expl}, the minimax criterion, and the I-optimal criterion with and without a split plot structure.
For space filling designs, the maximin criterion is often used, which is reported as the minimum $L_2$-distance between all points in the design (\cite{JohMooYlv90Mini}),
\begin{equation}
\min_{k\neq l}\|x_k - x_l\|_2,
\end{equation}
which is to be maximized. The rational behind this criterion is to avoid situations with two runs being too close to each other, as this does not provide much more information.
Related to this, \cite{MorMit95expl} suggested an alternative, which does only consider the minimum distance:
\begin{equation}
\Phi_p(D) = \sum_{j = 2}^{n_{overall}}\sum_{l = 1}^{j - 1}\|x_j - x_l\|^{-p},
\end{equation}
with $p = 2$ here. The criterion $\Phi$ needs to be minimized.
Another characterization of space filling properties is done by minimax designs, where one is interested in the distance of an arbitrary point $x$ in the design space to the design points $x_j, j = 1, \dots, n_{overall}$:
\begin{equation}
d(x) = \min_{j = 1, \dots, n_{overall}}\|x - x_j\|
\end{equation}
The main rational behind minimax designs is to minimize the maximum value for $d(x)$, denoted by $mM$. This implies, that for each point in the design space, a design point is at maximum $mM$ away. 
As the minimax criterion is hard to calculate directly, a Monte Carlo estimate is used.

On the other hand, typical design criteria in linear models are D and I. While D-optimal designs aim to estimate the parameters of a linear model optimal in the sense of the covariance of the least squares estimator, I-optimal designs target a minimal average variance of the least squares prediction over the design space. This is why we chose I-optimality. Both of these heavily depend on the Fisher Information matrix as well as on the underlying model assumption. Here, a quadratic model including all 2-factor interactions is assumed, i.e. a standard response surface model. The I-optimality criterion is defined as:
\begin{equation}
I(X) = 2^{-d}\int_{x \in [-1, 1]^d} f(x) (X' I X )f'(x)dx,
\end{equation}
with $I$ being the identity matrix representing the assumed covariance structure.
In order to adept for the error structure in split plot designs, the criteria needs to be updated with a different covariance matrix:
\begin{equation}
I(X) = 2^{-d}\int_{x \in [-1, 1]^d} f(x) (X' V X )f'(x)dx,
\end{equation}
where $V$ is a blockwise covariance matrix, representing the assumed error structure for equation (1). Please see  \cite{Goos2002optBlockedSplitPlotDoEBook} and \cite{JonesGoos2012IoptVsDopt} for more details and equations on I-optimality. 

\begin{figure}
	\centering
	\begin{subfigure}{.5\textwidth}
		\centering
		\includegraphics[width=.95\linewidth]{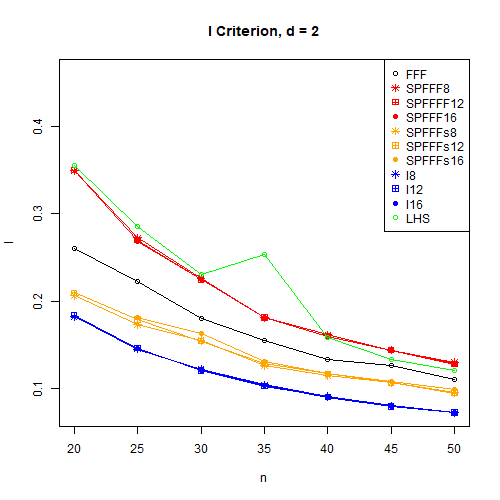}
	\end{subfigure}%
	\begin{subfigure}{.5\textwidth}
		\centering
		\includegraphics[width=.95\linewidth]{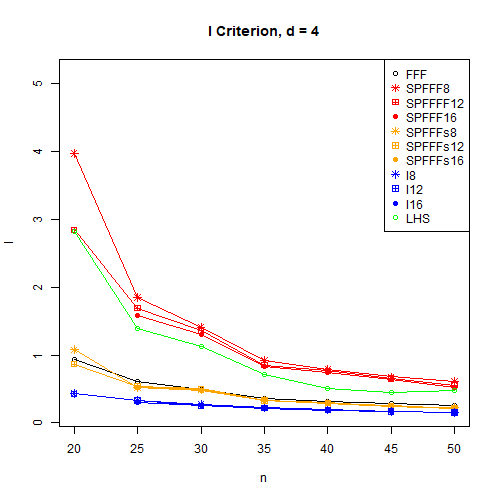}
	\end{subfigure}
	\caption{Comparison of I-optimality with 2 (left) and 4 (right) dimensions (smaller is better).}
	\label{fig:Iopt}
\end{figure}

\begin{figure}
	\centering
	\begin{subfigure}{.5\textwidth}
		\centering
		\includegraphics[width=.95\linewidth]{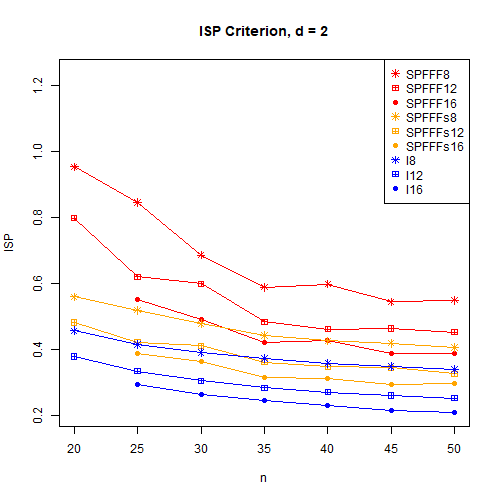}
	\end{subfigure}%
	\begin{subfigure}{.5\textwidth}
		\centering
		\includegraphics[width=.95\linewidth]{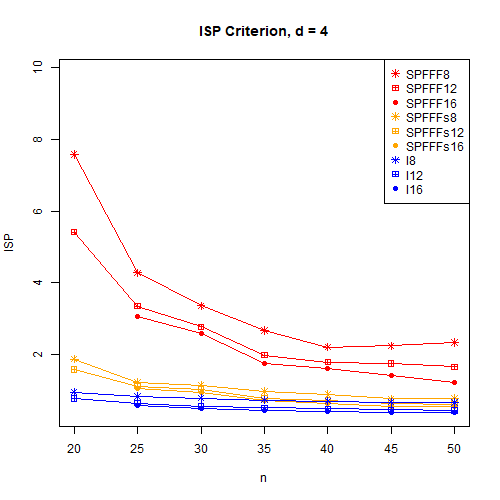}
	\end{subfigure}
	\caption{Comparison of I-optimality for Split Plot designs with 2 (left) and 4 (right) dimensions, assuming a variance ratio of 1 (smaller is better).}
	\label{fig:ISPopt}
\end{figure}

\begin{figure}
	\centering
	\begin{subfigure}{.5\textwidth}
		\centering
		\includegraphics[width=.95\linewidth]{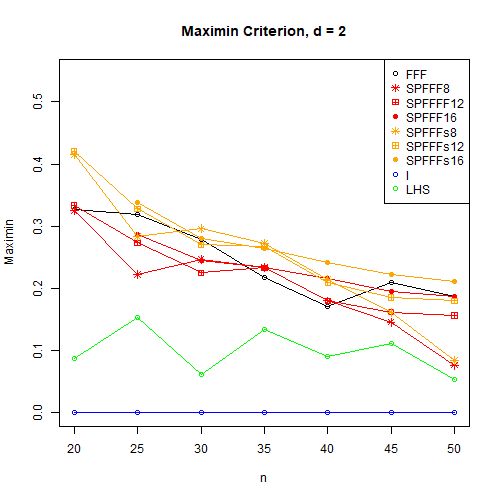}
	\end{subfigure}%
	\begin{subfigure}{.5\textwidth}
		\centering
		\includegraphics[width=.95\linewidth]{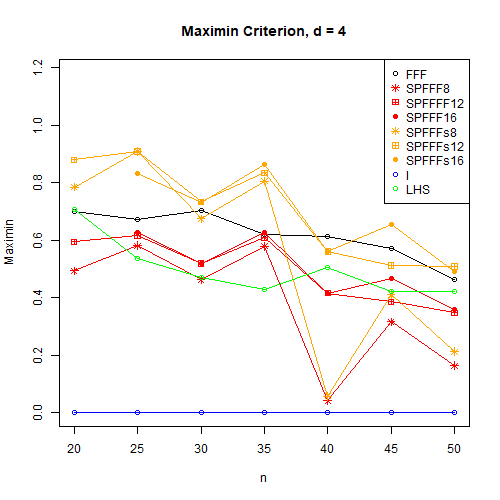}
	\end{subfigure}
	\caption{Comparison of Maximin criterion with 2 (left) and 4 (right) dimensions (larger is better).}
	\label{fig:MaximinOpt}
\end{figure}

\begin{figure}
	\centering
	\begin{subfigure}{.5\textwidth}
		\centering
		\includegraphics[width=.95\linewidth]{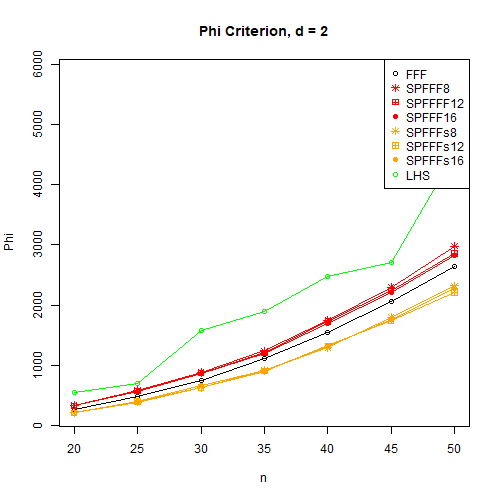}
	\end{subfigure}%
	\begin{subfigure}{.5\textwidth}
		\centering
		\includegraphics[width=.95\linewidth]{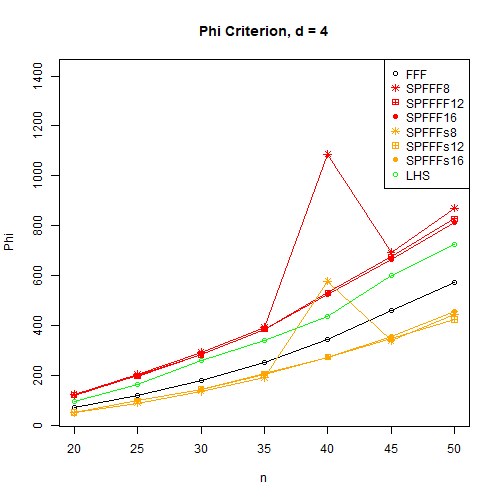}
	\end{subfigure}
	\caption{Comparison of $\Phi$-criterion with 2 (left) and 4 (right) dimensions (smaller is better).}
	\label{fig:PhiOpt}
\end{figure}

\begin{figure}
	\centering
	\begin{subfigure}{.5\textwidth}
		\centering
		\includegraphics[width=.95\linewidth]{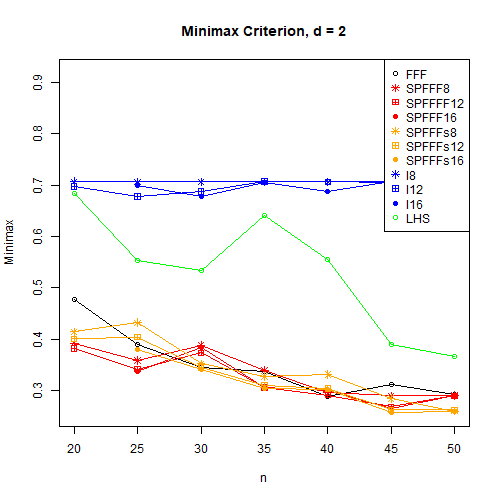}
	\end{subfigure}%
	\begin{subfigure}{.5\textwidth}
		\centering
		\includegraphics[width=.95\linewidth]{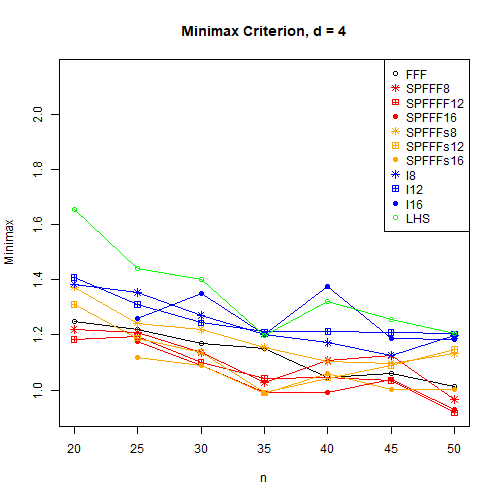}
	\end{subfigure}
	\caption{Comparison of Minimax criterion with 2 (left) and 4 (right) dimensions (smaller is better).}
	\label{fig:MinimaxOpt}
\end{figure}

The results of the comparison can be summarized as follows: for I-optimality and I-optimality including a split plot structure, the I-optimal designs are clearly preferable with a loss in efficiency to the next best alternative of around 50\%. Surprisingly, for I-optimality including a split plot structure, the scaled Split Plot FFF for larger $n$ are not out of reach.

For the Maximin criterion and the $\Phi_2$ criterion, the I-optimal results are not shown, as they have (by design) repeated points, resulting in either $0$ or $\inf$.

Analyzing Figure 1 to 5, several learnings are made:
\begin{itemize}
	\item For $I$-optimality, both including and excluding a split plot structure, the designs optimized for these criteria are with no doubt ideal under the corresponding assumptions. However, surprisingly, the scaled SPFFF designs are both for a iid error assumption as well as a blocking structure not out of reach compared to I-optimal designs. 
	\item For the Maximin and the $\Phi_2$ criterion, the SPFFF designs outperforms the others, surprisingly even better than the FFF designs.
	\item For the Minimax criterion, the FFF design, the unscaled and scaled SPFFF designs are on a comparable level. 
	\item The different numbers of whole plots (blocks) (8, 12 or 16) do not look like having a large impact on optimality besides for the I-optimality considering the blocking structure. 
\end{itemize}
Hence, it can be concluded, that if there is a need for introducing a blocking structure, in most cases, there will no strong loss of design efficiency. The scaled SPFFF is preferred in all cases over the unscaled one, meaning that the resulting DoE of the presented algorithm should be scaled to have design points on the boundary of the design space.

\subsection{Prediction performance comparison}
In this section, the following 4 methods will be applied to 2 analytical examples for a small comparison of prediction power in different designs.
\begin{table}
\begin{tabular}{L{3cm}|c|c|C{2cm}|C{6cm}}
Method & Continuous & Categorical & R-package & Details\\
\hline	
generalized linear models & yes & yes & base-R (glm) & Quadratic model\\
\hline 
Kriging & yes & no & DiceKriging (km) & constant regression term, gaussian cor. function, ML estimator\\
\hline 
Support vector machines & yes & yes & e1071 & defaults as in command \verb|svm| \\
\hline
Neural Network & yes & yes & keras (tensorflow) & 2 fully connected, hidden layers with 5 nodes each. Relu activation function, MSE or Cross Entropy loss 
\end{tabular}
\caption{Models used for prediction performance comparison.}
\end{table}
As models, the following options are taken. Details are described below and are listed in Table 2.
	\begin{itemize}
		\item Standard Least Squares and Generalized Least Squares as described in the command "glm" (\cite{R2019}) using a model with main effects and all 2-factor interactions. 
		\item Kriging implemented in the package DiceKriging, (\cite{DiceKriging}) using a constant regression term and Matern $5/2$ correlation function.
		\item Support vector machines as implemented in package e1071 (\cite{R_e1071}).
		\item Feedforward Neural Networks as implemented in the the package keras using tensorflow backend (\cite{KerasR}) with 
		2 hidden layers each having 5 nodes. The Relu activation function is used and for the categorical output a sigmoid function as last layer and for the regression problem a linear activation function. As loss functions, cross entropy and mean square error loss have been used.
	\end{itemize}
 As test beds, two examples are used. To reflect the situation of the motivating example, no random noise is added to the models. The cantilever example is described on the following website \verb|http://www.sfu.ca/~ssurjano/index.html| (\cite{simulationlib}) and is used on several simulation studies. The part used here represents the displacement of a cantilever beam. In order to have a pass/fail output as well, a sigmoid function is applied to the numeric output and rounded to 0/1 (here indicated by the $[.]$ brackets).
\begin{equation}
D(x) = \frac{4L^3}{Ewt}\sqrt{\left( \frac{Y}{t^2}\right)^2 + \left(\frac{X}{w^2}\right)^2}
\end{equation}
\begin{equation}
PF(x) = \left[\text{sigmoid}\left(\frac{4L^3}{Ewt}\sqrt{\left( \frac{Y}{t^2}\right)^2 + \left(\frac{X}{w^2}\right)^2} - 4.3\right)\right]
\end{equation}

with $L=100$, $D_0 = 2.2535$. $w$ and $t$ representing the width and thickness of the cross-section of the beam are set to $4$ and $2$ respectively. The inputs are $R \in [36000, 44000], E \in [2.61e7, 3.19e7], X \in [300, 700]$ and $Y\in[800,1200]$.\\

Being just one simulation example, it indicates already how the prediction performance does not depend so much on the design type but rather on the applied model and the sample size. Although this should not be understood as an argument to take less effort to plan an experiment (all used designs here are already optimized in different ways) but there is some flexibility in how to design an experiment and to tailor designs such that they fit to specific project requirements.

\begin{figure}
\centering	\includegraphics[width=.95\linewidth]{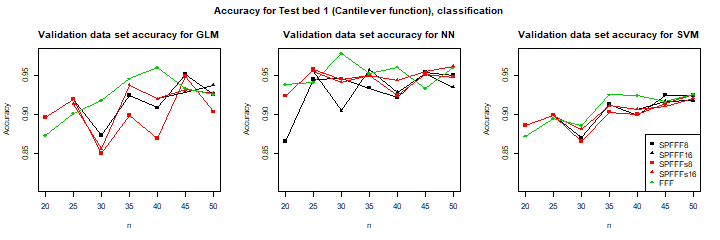}
\caption{Comparison of prediction power for classification. A Validation data set with 100k uniformly spread over the design space has been used to calculate the accuracies.}
\label{fig:ClassPredPowerComp}
\end{figure}

\begin{figure}
	\centering	\includegraphics[width=.95\linewidth]{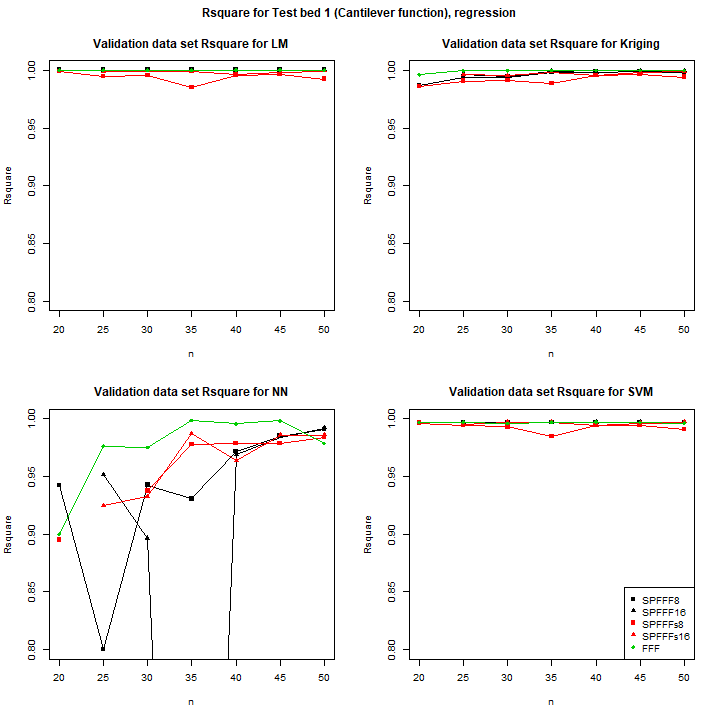}
	\caption{Comparison of prediction power for regression. A Validation data set with 100k uniformly spread over the design space has been used to calculate the accuracies.}
	\label{fig:RegPredPowerComp}
\end{figure}

\section{Summary}

In this paper, a modification of a spacefilling strategy is suggested and evaluated, which incorporates a split plot-like blocking structure into the factors. This reflects a challenge that occurs in industrial applications, when not all factors are equally easy to change. Comparing the performance to other designs under different optimality criteria, it turns out that the designs do not loose performance against other space filling criteria. Furthermore, in some situations it still performs acceptable with respect to I-optimality in some situations. Yet when applying these designs in a simulated, deterministic emulation example, performance against other space filling designs was on a similar level. Some next steps could be to also incorporate qualitative factors, or factor constraints into the designs, as mentioned in the introduction. 

\bibliographystyle{apalike}

\end{document}